\documentclass[%
preprint,
 amsmath,amssymb,
 aps,
prb,
 longbibliography,
 lengthcheck,%
]{revtex4-1}

\usepackage{graphicx}
\usepackage{dcolumn}
\usepackage{bm}
\usepackage{hyperref}

\begin{document}


\title{Visualizing the Pt Doping Effect on Surface and Electronic Structure 
in Ir$_{\mathrm{1-x}}$Pt$_{\mathrm{x}}$Te$_{\mathrm{2}}$ by Scanning Tunneling Microscopy and Spectroscopy}

\author{Y. Fujisawa$^{1}$}
\email{1213625@ed.tus.ac.jp}
\author{T. Machida$^{1,2}$, K. Igarashi$^{1}$, A. Kaneko$^{1}$, T. Mochiku$^{2}$, S. Ooi$^{2}$, M. Tachiki$^{2}$, K. Komori$^{2}$, K. Hirata$^{2}$}

\author{H. Sakata$^{1}$}

\affiliation{\\$^{1}$Department of Physics, Tokyo University of Science, 1-3 Kagurazaka, Shinjuku-ku, Tokyo 162-8601, Japan \\$^{2}$Superconducting Materials Center, National Institute for Materials Science, 1-2-1 Sengen, Tsukuba, Ibaraki 305-0047, Japan}
\date{\today}

\begin{abstract}
We report on the Pt doping  effect on surface and electronic 
structure in Ir$_{\mathrm{1-x}}$Pt$_{\mathrm{x}}$Te$_
{\mathrm{2}}$ by scanning tunneling microscopy (STM) and spectroscopy (STS).
The surface prepared by cleavage at 4.2 K shows a triangular lattice of 
topmost Te atoms.
The compounds that undergo structural transition have supermodulation 
with a fixed wave vector $q = \frac{2\pi}{5a_m}$ (where $a_m$ is the lattice constant in 
the monoclinic phase) despite the different Pt concentrations.
The superconducting compounds show patch structures.
The surface of the compound that exhibits neither the superconductivity nor 
the structural transition shows no
superstructure.
In all doped samples, the dopant is observed as a dark spot in STM images.
The tunneling spectra near the dopant show the change in the local density of 
state at approximately -200 mV.
Such microscopic effects of the dopant give us the keys for establishing a 
microscopic model of this material.
\end{abstract}

\pacs{Valid PACS appear here}
\maketitle
In transition metal compounds with $d$ electrons, various exotic 
phenomena are expected because of strong
correlation of electrons.
One of the most intriguing phenomena is high-$T_{\mathrm{c}}$ 
superconductivity (SC) in cuprates and iron-based
superconductors, which have  antiferromagnetic and/or orbital orders in 
their parent materials.
Ir compounds, which also have $d$ electrons, have been attracting much 
attention.
Exotic phenomena such as high-$T_{\mathrm{c}}$ SC are expected owing to the 
competition between the electronic correlation
and the spin-orbit interaction\cite{Wang_1}.
Recently, IrTe$_{\mathrm{2}}$ has become a new member of the Ir compound 
family.
This material undergoes a structural transition from trigonal to monoclinic 
(triclinic) structure at about 280 K
accompanied by the abrupt increase in electric resistivity.
In the high-temperature phase, the edge-sharing IrTe$_{\mathrm{6}}$ octahedra form a layered structure with an Ir equilateral lattice along the c-axis, as shown in Figs. 1(a) and 1(b). In the low-temperature phase, the Ir lattice changes into an isosceles triangular lattice, and unit cell vectors are also changed to $a_m$ and $b_m$, as shown in Fig. 1(c).
With chemical doping (such as Pt, Pd, and Rh) at the Ir site, the structural 
transition is suppressed and the SC appears
at around 3 K\cite{Pyon_1, Yang_1, Kudo_1}.
As in the case of iron-based superconductors, it is expected that an ordered 
state exists below the structural
transition temperature ($T_{\mathrm{s}}$) in  IrTe$_{\mathrm{2}}$, and the 
emergence of the SC in  doped IrTe$_{\mathrm
{2}}$ is related to the disappearance of the ordered state.
In IrTe$_{\mathrm{2}}$, the existence of supermodulation (SM) with the wave vector
$q = \frac{2\pi}{5a_m}$  (where $a_m$ is the lattice constant in 
the monoclinic phase) has been reported from 
diffraction\cite
{Seok_1, Yang_1, Cao_1, Pascut_1, Toriyama_1} and STM\cite{Machida_1} 
studies.
This SM may relate to the expected ordered state.
Therefore, it is interesting to investigate how such SM evolves with 
chemical doping and how the dopant affects the
structure and electronic states. To answer these questions, we have performed scanning tunneling 
microscopy (STM) and spectroscopy (STS) on Ir$_{\mathrm{1-x}}$Pt$_{\mathrm{x}}$Te$_{\mathrm{2}}$.
We observed SM in the low-temperature phase and patchwork structures in the superconducting sample.
Furthermore, Pt sites were identified by STM measurements.
The STS measurements revealed the change in the local density of states 
(LDOS) around -200 mV near the dopant site.

The single crystals used in this study were grown by the self-flux method.
In order to determine $T_{\mathrm{s}}$ and the superconducting transition 
temperature ($T_{\mathrm{c}}$), we measured the temperature dependence of
electric resistivity.
The normalized resistivity is shown in Fig. 1(d).
The resistivities of the  compounds with $x=0.00, 0.02$, and 0.04 show a sudden 
increase, suggesting the existence of the structural transition.
The sample with $x=0.07$ shows superconducting transition at 2.8 K [Fig. 1(e)].
The compound with $x = 0.15$ shows no superconducting transition above 2 K.
The structural and superconducting transition temperatures are 
summarized in Table I.
These results are qualitatively the same as those reported 
previously\cite{Pyon_2}.
However, the reported Pt concentration at which $T_{\mathrm{c}}$ appears is much lower 
than our nominal value.
Furthermore, the same structural transition temperatures ($T_{\mathrm{s}}$) appear at much 
higher concentrations in our samples than in reported samples.  
This indicates that, in our single crystals, the actual concentrations of the 
dopant are expected to be lower than the nominal ones.
We used a laboratory-made scanning tunneling microscope for STM and STS 
measurements.
All the measurements were carried out at 4.2 K in He gas.
Because of the weak van der Waals coupling between the two Te layers, a 
clean sample surface is obtained by cleavage at
4.2 K.
An electrochemically polished Au wire or a mechanically cut Pt-Ir wire (Pt:Ir = 
80\%:20\%) was used for the STM measurements.
A difference in the tips did not affect the results.
STM images are obtained in the constant-current mode.
The differential tunneling conductance ($dI/dV$) was obtained by the numerical 
differentiation of the $I-V$ characteristics.

\begin{table*}
\label{t1}

\begin{center}

\begin{tabular}{lclclclclc}

\hline
\multicolumn{1}{c}{\shortstack{Nominal concentration\\ of dopant}} & 
\multicolumn{2}{c}{\shortstack{Concentration of
dark spots}}& \multicolumn{2}{c}{\shortstack{$T_s$ (K)}} & \multicolumn{2}
{c}{\shortstack{$T_c$ (K)}}& \multicolumn{2}{c}{\shortstack
{Superstructure}}\\
\hline
\multicolumn{1}{c}{\shortstack{0.00}} & \multicolumn{2}{c}{\shortstack
{0.000}} & \multicolumn{2}{c}{\shortstack{280}} &
\multicolumn{2}{c}{\shortstack{-}}&
\multicolumn{2}{c}{\shortstack{SM}}\\
\hline
\multicolumn{1}{c}{\shortstack{0.02}} & \multicolumn{2}{c}{\shortstack
{0.008}}& \multicolumn{2}{c}{\shortstack{218}} &
\multicolumn{2}{c}{\shortstack{-}}&
\multicolumn{2}{c}{\shortstack{SM}}\\
\hline
\multicolumn{1}{c}{\shortstack{0.04}} & \multicolumn{2}{c}{\shortstack
{0.01}}& \multicolumn{2}{c}{\shortstack{212}} &
\multicolumn{2}{c}{\shortstack{-}}&
\multicolumn{2}{c}{\shortstack{SM}}\\
\hline
\multicolumn{1}{c}{\shortstack{0.07}} & \multicolumn{2}{c}{\shortstack{0.025}}& \multicolumn{2}{c}{\shortstack{-}} &
\multicolumn{2}{c}{\shortstack{2.8}}&
\multicolumn{2}{c}{\shortstack{PS}}\\
\hline
\multicolumn{1}{c}{\shortstack{0.15}} & \multicolumn{2}{c}{\shortstack
{0.073}}& \multicolumn{2}{c}{\shortstack{-}} &
\multicolumn{2}{c}{\shortstack{-}}&
\multicolumn{2}{c}{\shortstack{-}}\\
\hline
\hline
\end{tabular}
\end{center}
\caption{List of samples used in this study. The concentration of the dark 
spots is estimated from the STM images shown in
Figs. 2(a)-2(e).
$T_{\mathrm{s}}$ and $T_{\mathrm{c}}$ are estimated from the electric 
resistivity measurements of each sample.}
\label{Fig_Gmap}

\end{table*}

\begin{figure}[tbh]
\begin{center}
\includegraphics[width=8cm]{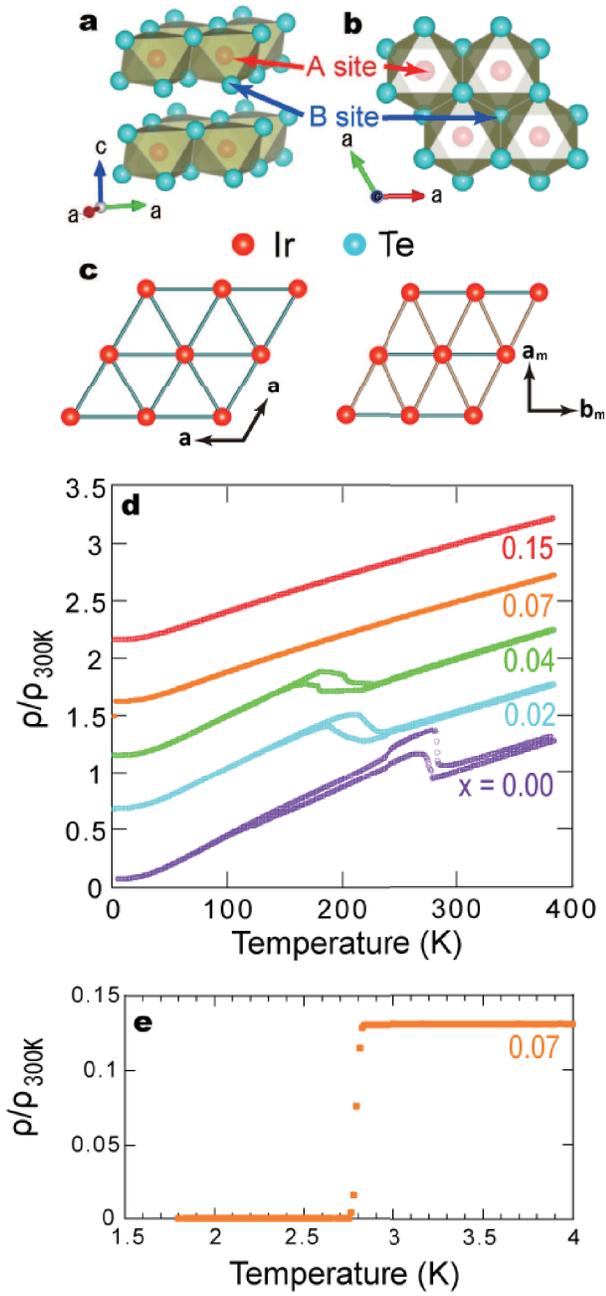}
\end{center}
\caption{(color online) (a) and (b), Crystal structure of IrTe$_{\mathrm
{2}}$ visualized using VESTA.
(c), Ir equilateral triangular lattice in high-temperature phase and isosceles triangular lattice in low-temperature phase.
(d), Temperature dependence of normalized resistivity $\rho/\rho_{\mathrm{300 
K}}$ of  Ir$_{\mathrm{1-x}}$Pt$_{\mathrm
{x}}$Te$_{\mathrm{2}}$ ($x = 0.00, 0.02, 0.04, 0.07$, and 0.15).
Each $\rho/\rho_{\mathrm{300 K}}$ is shifted upwards by 0.5 for clarity.
(e), $\rho/\rho_{\mathrm{300 K}}$ of Ir$_{\mathrm{0.93}}$Pt$_{\mathrm{0.07}}$Te$_{\mathrm{2}}$ below 4 K.}
\label{Fig_Gmap}
\end{figure}

\begin{figure*}[tb]
\begin{center}
\includegraphics[width=14cm]{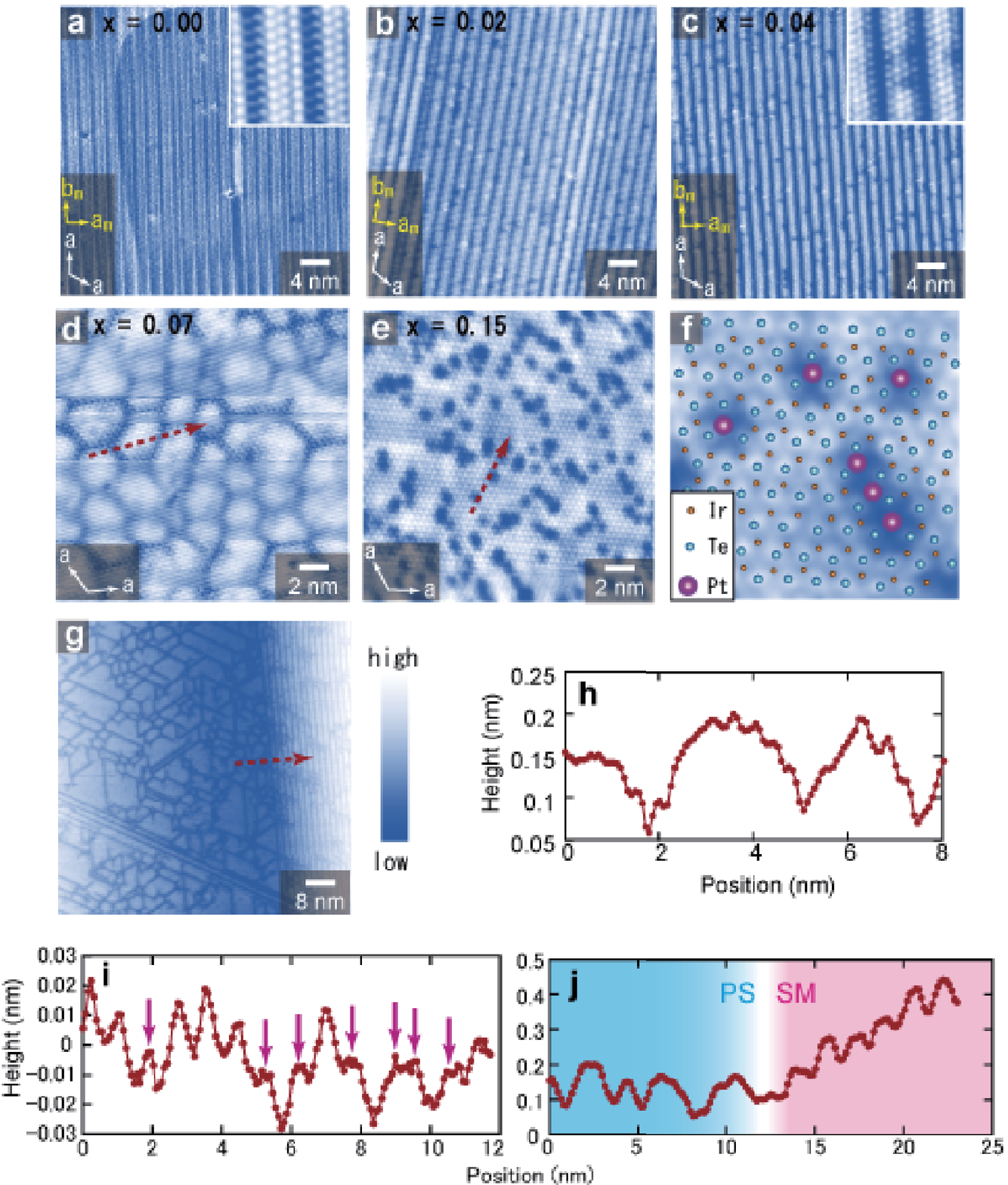}
\end{center}
\caption{(Color online) (a) to (e), Typical STM images of  Ir$_{\mathrm{1-
x}}$Pt$_{\mathrm{x}}$Te$_{\mathrm{2}}$ ($x =
0.00, 0.02, 0.04, 0.07$, and 0.15).
Set bias is 500 mV.
In all the compounds, the surface is composed of a triangular lattice of topmost 
Te atoms.
SM with a wave vector $q = \frac{2\pi}{5a_m}$ can be observed in Figs. 2(a)-2(c).
In Fig. 2(d), Te atoms form patches with an area of a few nm$^{\mathrm{2}}$, and 
these patches are randomly paved.
In Fig. 2(e), superstructures such as SM and PS cannot be seen.
(f), Magnified STM image of Fig. 2(e).
The dark area spreads over the neighboring three Te atoms and the darkest 
position is the center of the three Te atoms.
(g), STM image that shows the coexistence of the SM and the PS in a sample with $x=0.02$.
(h), STM line profile along the red broken arrow in Fig. 2(d). The height variation of the boundary of patches is about 0.1 nm.
(i), STM line profile along the red broken arrow in Fig. 2(e). Purple 
arrows indicate the Te atoms that are depressed by the dopant Pt. The heights of 
these Te atoms are about 0.02 nm lower than those of the others.
(j), STM line profile along the red broken arrow in Fig. 2(g). The height 
of the patch is almost the same order as the amplitude of the SM.
}
\end{figure*}

Figures 2(a)-2(e) show typical STM images on Ir$_{\mathrm{1-x}}$Pt$_
{\mathrm{x}}$Te$_{\mathrm{2}}$ for $x=0.00, 0.02,
0.04, 0.07$, and $0.15$, respectively.
The triangular lattice composed of topmost Te atoms can be seen in all 
images.
In addition to the atomic lattice, the SM with $q = \frac{2\pi}{5a_m}$, 
 which is the same as that reported previously\cite{Machida_1},
exists in the compounds with $x=0.00$, as shown in Fig. 2(a).
The SM with the same modulation wave vector is also observed in the 
compounds with $x = 0.02$ and 0.04 [Figs. 2(b) and 2(c)].
Thus, the period of the modulation does not change in the low-temperature 
phase despite the different amounts of the
dopant.
However, in the compound with $x=0.07$, which shows SC, we found that 
patches with areas of a few nm$^{\mathrm{2}}$ are
formed.
The surface is randomly paved with patches, as shown in Fig. 2(d).
We call this peculiar structure ``patchwork structure (PS)''.
The boundary of the patches tends to be perpendicular to the Ir-Te 
bond directions.
Note that the SM in the low-temperature phase is also perpendicular to the Ir-Te bonds.
The size of the patches is random.
Furthermore, in the compound with $x = 0.15$, which shows neither structural transition nor SC, no superstructure appears, as shown in Fig. 2(e). 
Thus, the surface structure seems to depend on the ground state of the material.
The coexistence of the SM and the PS was seen in a part of 
the sample, which is perhaps due to the  non uniform distribution of the 
dopant. An example in a sample with $x = 0.02$ is shown in Fig. 2(g). The change in the surface from the SM to the PS is rather abrupt and we have not observed other types of surface 
structure, such as the SM with a different modulation period. The size 
of the patch observed in such a region is larger than that observed in Ir$_{0.93}$Pt$_{0.07}$Te$_{\mathrm{2}}$ shown in Fig. 2(d).

As can be seen from Figs. 2(b)-2(e), there are several dark spots in STM 
images on Pt-doped samples.
This is easy to see in Fig. 2(e) because of the lack of the SM and the PS.
The magnified STM image, shown in Fig. 2(f), reveals that the dark region, 
which is seen as the dark spots in Fig. 2(e),
spreads over the neighboring three Te atoms and the center of the three Te atoms 
is darkest.
This darkest position can be the Ir site (we refer to this site as the A site) or the Te 
site, which is in the layer below the topmost
Te layer (we refer to this site as the B site), as shown in Figs. 1(a) and 1(b).
The A and B sites cannot be distinguished by STM measurements, because both 
sites are located at the center of neighboring the three Te atoms in STM images.
However, as can be seen from the STM images in Fig. 2(f), 
all the darkest positions are connected by a linear combination of the unit 
vector $\boldmath{a}$.
This means that all such sites are located in either the A or B site.
We found that the number of dark spots changed with Pt doping.
We measured the number by counting the dark spots in the STM images.
The concentrations of the dark spots are summarized in Table I.
Because the concentration of dark spots increases with Pt doping, and all the 
darkest positions can be A sites, it is plausible that each darkest position 
corresponds to the Pt site doped at the Ir site.
Note that the concentrations of the dark spots are 
less than the nominal dopant concentrations.
This is because, as we mentioned, the actual concentrations in our samples are expected to be less than 
the nominal ones from the resistivity measurements.
Figures 2(h)-2(j) show the cross-sectional profiles of each STM image.
These indicate height variations in STM images due to the SM, the boundary of 
patches, and the dopant.
As can be seen from these profiles, the height variation of both the SM and 
the boundary of patches is about 0.1 nm.
On the other hand, the variation due to the dopant is about 0.02 nm.

\begin{figure}[tb]
\begin{center}
\includegraphics[width=8cm]{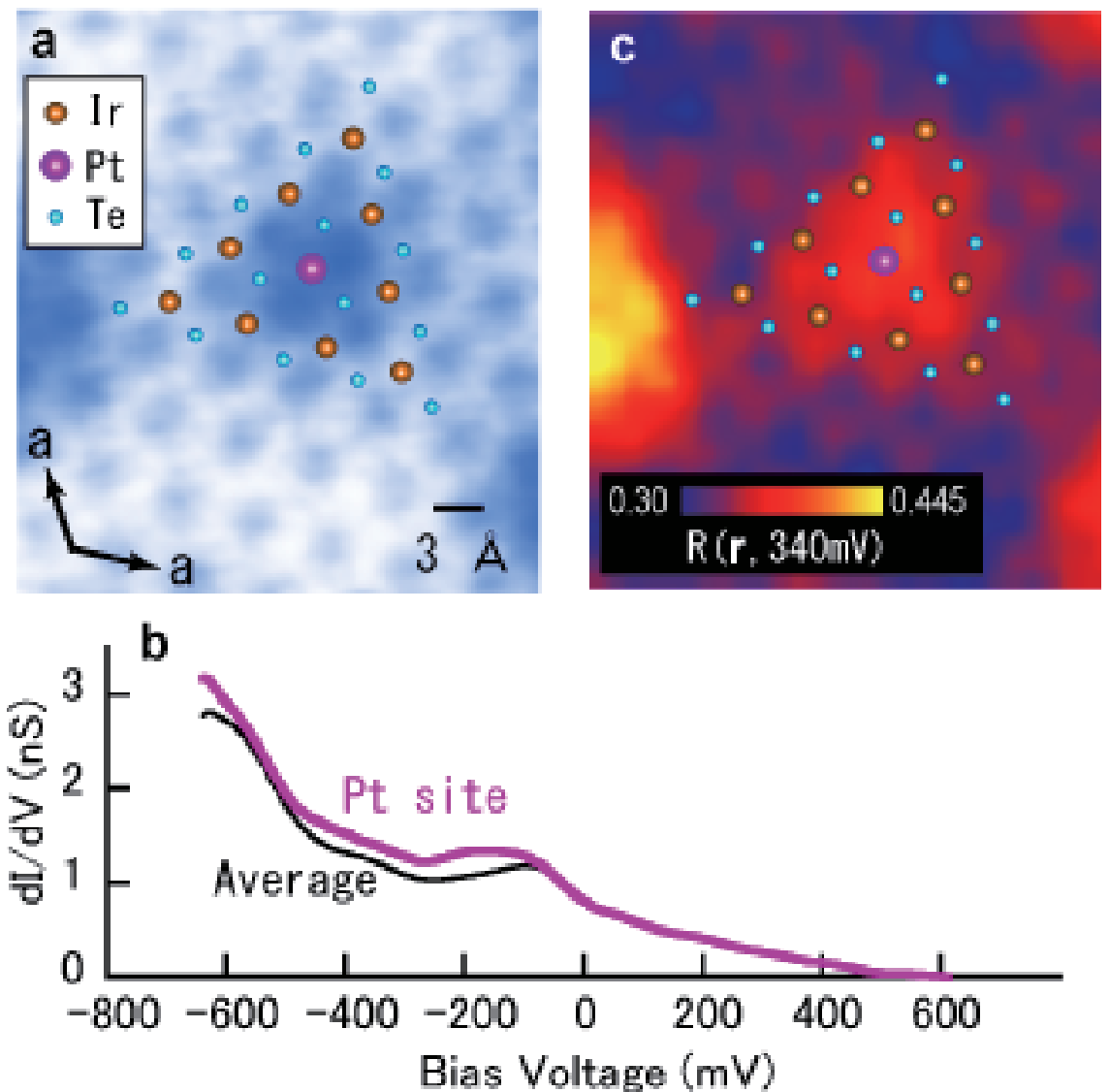}
\end{center}
\caption{(Color online) (a), An STM image that includes a dopant site taken 
at $V_{\mathrm{set}}/I_{\mathrm{set}}=500
\mathrm{mV}/500 \mathrm{pA}$.
(b), Spectrum taken in the dark region (purple) and averaged spectrum in the other region (black).
Set point is $V_{\mathrm{set}}$/$I_{\mathrm{set}}$ = 500 mV/500 pA.
(c),  Map of the ratio of the tunneling currents at -340 mV to that 340 mV; $R(\mathrm{\textbf{r}}, \mathrm{340 mV})$
= $I(\mathrm{\textbf{r}}, \mathrm{-340 mV})$/$I(\mathrm{\textbf{r}}, 
\mathrm{+340 mV})$.}
\label{fig3}
\end{figure}

We also observed the dopant effect on the LDOS by STS measurements.
As previously reported, the LDOS is modulated by the SM\cite{Machida_1}.
Thus, such a structure makes the interpretation of the LDOS change by dopants 
difficult.
Therefore, we examined the LDOS change due to the dopant, which 
is located on a patch in Ir$_{\mathrm{0.93}}$Pt$_{\mathrm{0.07}}$Te$_{\mathrm{2}}$. 
Figure 3(a) shows an STM image with a dark region, which includes the dopant 
site.
The set bias voltage was 500 mV.
The averaged spectrum taken in the dark region (purple) and that in the other region (black) are shown in Fig.
3(b).
Both spectra exhibit energy asymmetry, as previously reported \cite
{Machida_1}.
Whereas there is no difference between them above $E_{\mathrm{F}}$, the 
enhancement of
$dI/dV$ appears at approximately -200 mV in the dark region, as can be seen in Fig. 3(b).
Because $dI/dV$ is proportional to the LDOS, this enhancement of $dI/dV$ is 
related to the change in the LDOS.
Figure 3(c) shows a map of the ratio of the tunneling current at -340 mV 
to that at 340 mV; $R(\textbf{r}, \mathrm{340
mV}) = I(\textbf{r}, \mathrm{-340  mV})/I(\textbf{r}, \mathrm{+340 mV})$.
This value indicates the ratio of the integrated LDOS between -340 meV and $E_{\mathrm
{F}}$ to that between $E_{\mathrm{F}}$ and  340 meV.
As can be seen from the figure, the brighter region, which indicates the 
enhanced LDOS ratio, corresponds to the region where the STM image shows the contrast.
The contrast of the STM image comes from both the 
topographic variation due to the dopant and the LDOS contributions.
Thus, the correspondence between the enhancement of LDOS and the STM 
contrast possibly indicates that the LDOS change is related to the local 
deformation of the lattice.

We found that the wave vector of the SM does not change in the low-temperature phase despite the increase in Pt concentration.
Thus, this wave vector is considered to be inherent in the low-temperature 
phase.
On the other hand, PS changes its average size with Pt concentration, as can be seen from Figs. 2(c) and 2(g).
PS may contribute to the reduction of the structural stress due to the Pt dopant.
However, the correlation between the appearance of PS and SC is an open 
question.

As for the electronic states, Ootsuki {\it et al.} reported the spectral 
weight suppression in occupied states in a pristine sample when the 
temperature is decreased below $T_{\mathrm{s}}$.
They also observed that the spectral weight suppression disappeared with Pt 
doping\cite{Ootsuki_1}.
Our observation of the change in LDOS near Pt sites is consistent with this 
result.

In conclusion, by STM, we have observed a triangular lattice composed of 
topmost Te atoms and various superimposed
structures of  Ir$_{\mathrm{1-x}}$Pt$_{\mathrm{x}}$Te$_{\mathrm{2}}$ ($x = 
0.00, 0.02, 0.04, 0.07$, and 0.15).
The SM with $q = \frac{2\pi}{5a_m}$ was observed in the 
compounds with $x = 0.00, 0.02$, and 0.04.
The PS was observed on the compound with $x=0.07$.
A modulated surface such as the SM or the PS vanished in the compound with $x = 
0.15$.
In all doped compounds, several dark spots due to the Pt dopants are observed.
STS measurements revealed that the LDOS at approximately -200 meV is changed by the 
dopant.
These local effects of Pt dopants on IrTe$_{\mathrm{2}}$ will provide 
considerable information for the establishment of a theoretical model of this system.

\end{document}